\documentclass{article}
\usepackage{spconfa4,amsmath, bm, amssymb, type1cm}

\newcommand{\argmin}{\mathop{\rm arg~min}\limits}
\usepackage{graphicx}
\usepackage{color}
\title{Monaural source enhancement maximizing source-to-distortion ratio via automatic differentiation}
\name{Hiroaki Nakajima, Yu Takahashi, Kazunobu Kondo and Yuji Hisaminato}
\address{Yamaha Corporation}
\begin{document}
%
\maketitle
\begin{abstract}
Recently, deep neural network (DNN) has made a breakthrough in monaural source enhancement. Through a training step by using a large amount of data, DNN estimates a mapping between mixed signals and clean signals. At this time, we use an objective function that numerically expresses the quality of a mapping by DNN. In the conventional methods, L1 norm, L2 norm, and Itakura-Saito divergence are often used as objective functions. Recently, an objective function based on short-time objective intelligibility (STOI) has also been proposed. However, these functions only indicate similarity between the clean signal and the estimated signal by DNN. In other words, they do not show the quality of noise reduction or source enhancement. Motivated by the fact, this paper adopts signal-to-distortion ratio (SDR) as the objective function. Since SDR virtually shows signal-to-noise ratio (SNR), maximizing SDR solves the above problem. The experimental results revealed that the proposed method achieved better performance than the conventional methods.
\end{abstract}
\begin{keywords}
source enhancement, source-to-distortion ratio, and Deep Neural Network
\end{keywords}
\section{Introduction}
\label{sec:intro}

In current acoustic research, monaural source enhancement is one of the areas being actively studied. Its applications range from preprocessing of speech recognition to remixing, and so on. Many techniques have  made remarkable achievements in source enhancement, for example, non-negative matrix factorization~\cite{smaragdis2007supervised, kitamura2014music}, Bayesian methods~\cite{ozerov2007adaptation}, and an analysis of sound structure~\cite{ono2008separation}.\par
On the other hand, deep neural network (DNN) has achieved a successful outcome in many research field~\cite{hinton2012deep, simonyan2014very, oord2016wavenet, saito2018statistical}. It is because DNN can model a relationship between input and output more accurately and flexibly than conventional methods. \par
Taking this advantage, DNN-based source enhancement technologies have emerged as a powerful solution~\cite{hershey2016deep, rethage2017wavenet, koizumi2017dnn}. In general, a DNN-based source enhancement consists of two steps. First, they train DNN using an enormous pairs of a mixture signal and a clean signal. This means that DNN searches the optimal mapping, which expresses the relation between mixture signals and clean signals. At this time, an objective function showing goodness of the mapping is prepared. It should be carefully prepared. For example, L1 norm and L2 norm between a clean signal and an estimated signal are representative examples. We call this process as a training step hereafter. Second,  an enhanced signal is estimated from an unknown mixture signal by using the trained DNN. We call this process as a separation step in the following.\par
However, conventional DNN does not use an appropriate goodness measure to optimize the performance of noise reduction. Hereinafter, we are going to discuss typical three objective functions form this vewpoint. First, L1 norm and L2 norm are most general objective function in the field of DNN research. They are very simple and easy to handle, but they only indicate how close the clean signal and the estimated signal. In other words, L1 norm and L2 norm allow the noise signal to approximate the clean signal. This is not suitable for source enhancement. Second, Itakura-Saito divergence~\cite{fevotte2009nonnegative, kitamuraIDLMA} is also used in the time-frequency domain.
This is widely used and based on signal distribution. However, it cannot be said that the degree of signal separation is directly optimized since Itakura-Saito
divergence also measures similarity of the clean signal distribution and the estimated signal distribution. Third,~\cite{fu2017end, kolbaek2018monaural, Hui2018Training} use short-time objective intelligibility (STOI)~\cite{taal2011algorithm}, which is a kind of separation measurement. Nevertheless, STOI does not indicate degree of contamination. It only indicates how close the clean signal and the estimated signal are on the basis of human perception. In other words, it also allows the noise signal to approximate the clean signal.\par
In this paper, we propose a new DNN which maximize source-to-distortion ratio (SDR)~\cite{vincent2006performance} via automatic differentiation~\cite{baydin2015automatic} to suppress noise contamination directly. SDR is one of the widely-used and general-purpose evaluation score on source enhancement algorithms. SDR can be derived by linear operation, which means automatic differentiation is able to use in order to optimize DNN. Calculating SDR consists of two steps. First, using orthogonal projection, we decompose the estimated signal into two terms: a component of the clean signal, and a contaminated component. Second, we take a ratio of these two terms. In short, SDR takes signal-to-noise ratio (SNR) virtually. Therefore, maximizing SDR solves the problem of the conventional methods. Simple experiments are performed and reveal that proposed method outperforms DNN that use conventional objective functions.  
\section{Conventional methods}
\label{sec:conv}
As mentioned in Section 1, sound source enhancement using DNN consists of two steps: the training step and the separation step, respectively. First, we train DNN as 
\begin{align}
\Theta = \argmin_{\theta} \sum_{i}  L(X_{i}, \, Y_{i}; \, \theta),
\label{eq: 1}
\end{align}
where $X_{i}$ and $Y_{i}$ are input and correct data, $\theta$ is a parameter of DNN, and $L$ is the objective function, respectively. L2 and L1 norms are commonly used in the conventional methods, and they are defined as follows, respectively,
\begin{align}
L(X_{i}, \, Y_{i}; \, \theta) &= \bigl(f(X_{i}; \, \theta) - Y_{i} \bigl)^2, 
\\
L(X_{i}, \, Y_{i}; \, \theta) &= |f(X_{i}; \, \theta) - Y_{i}|, 
\end{align}
where $f(X_{i}; \, \theta)$ is a DNN's output, and $|\cdot|$ takes absolute. 
If $f(X_{i}; \, \theta)$ and $Y_{i}$ can be described as power spectrograms respectively, Itakura-Saito divergence is also widely used and defined as
\begin{align}
L(X_{i},\, Y_{i};\, \theta) = \sum_{k,m} \frac{y_{k,m}}{x_{k,m}} - \log \frac{y_{k,m}}{x_{k,m}}-1,
\end{align}
where $\bm{x}$ and $\bm{y}$ are $K \times M$ matrices, which represents $f(X_{i};\,  \theta)$ and $Y_{i}$, respectively. $x_{k,m}$ and $y_{k,m}$ are elements of $\bm{x}$ and $\bm{y}$. $K$ is the number of frequency bins, and $M$ is the number of time frames. Optimizing Itakura-Saito divergence is a maximum likelihood estimation assuming
$y_{k,m} \sim \exp (-y_{k,m}/x_{k,m})/x_{k,m}$.\par
Recently, STOI-based objective function has been proposed. It is defined as 
\begin{align}
L(X_{i},\, Y_{i};\, \theta) = \sum_{j,m} \frac{d_{j,m}}{JM},
\end{align}
where
\vspace{-5pt}
\begin{align}
d_{j,m} =& \frac{\bigl( \bar{y}_{j,m} - \mu(\bar{y}_{j,m})\bigl)^{T} \bigl( \hat{x}_{j,m}-\mu(\hat{x}_{j,m}) \bigl)}{ \|\bar{y}_{j,m} - \mu(\bar{y}_{j,m})\| \| \hat{x}_{j,m}-\mu(\hat{x}_{j,m})\|},\\
\hat{x}_{j,m} =& [\hat{\alpha}_{j, m-N+1},\, \hat{\alpha}_{j, m-N+2},\, ...,\, \hat{\alpha}_{j, m}]^T,\\
\hat{\alpha}_{j, l} =& \min \biggl(\frac{\| \bar{y}_{j,m}\|}{\|\bar{x}_{j,m}\|}\alpha_{j, l}, \, \bigl(1+10^{-\zeta/20} \bigl) \beta_{j, l} \biggl),\\
\bar{x}_{j,m} =& [\alpha_{j, m-N+1},\, \alpha_{j, m-N+2},\, ...,\, \alpha_{j,m}]^T,\\
\bar{y}_{j,m} =& [\beta_{j, m-N+1},\, \beta_{j, m-N+2},\, ...,\, \beta_{j, m}]^T,\\
\alpha_{j, m} =& \sqrt{\sum^{q_j-1}_{k=p_j } | x_{k,m}|^2},\\
\beta_{j, m} =& \sqrt{\sum^{q_j-1}_{k=p_j}  | y_{k,m}|^2},
\end{align}
$l \in \{m-N+1,\,...,\, m\}$, $N$ is an analysis length, $\zeta$ is the lower SDR bound, $\mu(\cdot)$ takes the sample average of the corresponding vector, $p_{j}$ and $q_{j}$ are the one-third octave band edges, and $J$ is the number of one-third octave bands corresponding to $j$, respectively. In summary, STOI-based objective function is regularized and weighted L2 norm according to human perception.\par
In a separation step, $\Theta$ is fixed, and unknown mixture data $Z$ is substituted as follows,
\begin{align}
C = f(Z; \, \Theta), 
\end{align}
where $C$ is enhanced data.

\section{Propose method}
\label{sec:propose}
\subsection{Overview of proposed method}
As mentioned in the previous section, conventional objective functions calculate similarity of estimated signal and clean signal. This is very natural good idea. However, the most important thing in source enhancement is to reduce the contaminated noise. Only emulating the clean signal through the noise signal is meaningless because conventional methods do not include any constraint for reducing the contaminated noise. Therefore, it is necessary to maximize noise reduction directly. \par
In this paper, we adopt SDR as the objective function in order to solve the aforementioned purpose. SDR shows SNR virtually via decomposing the
estimated signal into the clean-signal part plus the residual-error part and taking ratio of these two parts. Consequently, maximizing SDR means maximizing SNR approximately. 
\par
\begin{figure}
\begin{center}
\includegraphics[ width=8.4cm]{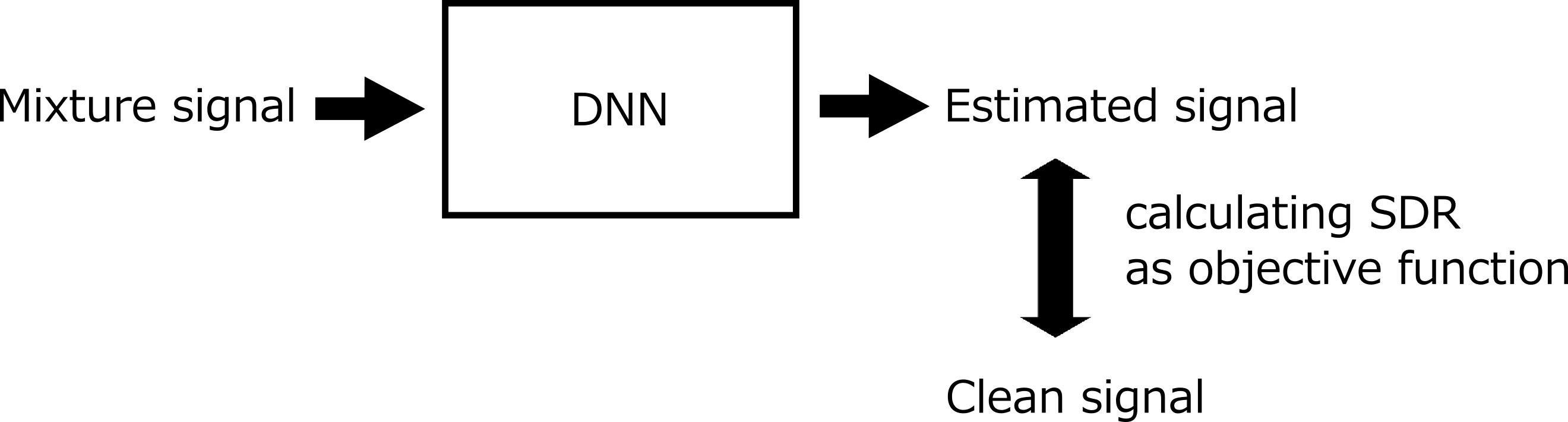}
\end{center}
\caption{Overview of proposed method.}
\label{fig: 1}
\vspace{-12pt}
\end{figure}

The proposed method in the training step is shown in Fig. \ref{fig: 1}. We use automatic differentiation to optimize DNN with SDR. Since SDR can be obtained only by linear operation, automatic differentiation is able to apply and this makes it easy to optimize DNN. 

\subsection{SDR}
SDR is one of the the evaluation score on source enhancement algorithms. It is defined as
\begin{align}
 {\rm SDR} = 10\log_{10} \frac{\| \bm{s}_{\rm{target}}{\|}^2}{\| \bm{\hat{s}} -\bm{s}_{\rm{target}}\|^2},
\end{align}
where $\| \cdot \| ^2$ takes power of signal. $\bm{\hat{s}}$ is the estimated signal described as $T\times 1$ vector, $T$ is the support of the signal, 
\begin{align}
\bm{s}_{\rm{target}} &= \bm{A}(\bm{A}^{T}\bm{A})^{-1}\bm{A}^T \bm{\hat{s}},
\label{eq: proj}
\end{align}
$\bm{A}$ is a $(T+G) \times G$ nonsymmetric Toeplitz matrix. Its first column is
\begin{align}
 A_{b, 1}&= \begin{cases}
    s_{b} & (1 \leq b \leq T), \\
    0 & ({\rm otherwise}),
\end{cases}
\end{align}
and first row is
\begin{align}
 A_{1, b}&= \begin{cases}
    s_{b} & (b=1), \\
    0 & ({\rm otherwise}),
\end{cases}
\end{align}
$G$ is the maximum delay allowed, $A_{d,e}$ and $s_{b}$ are elements of $\bm{A}$ and $\bm{s}$, respectively, and $\bm{s}$ is the clean signal described as $T\times 1$ vector. (\ref{eq: proj}) means that $\bm{s}_{\rm target}$ is a least squares solution of $\bm{w}$ in a following equation,
\begin{align}
 \bm{\hat{s}} = \bm{A}\bm{w},
\end{align}
where $\bm{w}$ is a $T \times 1$ vector. It means that $\bm{s}_{\rm target}$ is orthogonal projection onto the subspace spanned by $\bm{s}$ and delayed $\bm{s}$. In other words, $\|\bm{s}_{\rm target}\|^2$ corresponds to the amount of the clean signal in $\bm{\hat{s}}$, and $\|\bm{\hat{s}} - \bm{s}_{\rm target}\|^2$ corresponds to the amount of residual error (a sum of artificial distortion and a noise-signal part). Therefore, it is conceivable that SDR takes a ratio of reconstructing quality for the estimated signal and the amount of distortion using engineering approach. 

\subsection{Automatic differentiation}
Automatic differentiation is a technique to calculate partial derivation automatically. This technique is based on a simple idea that any function is composed of basic arithmetic operations such as addition, subtraction, multiplication, division, and basic functions (for example, exponential function, logarithmic function, trigonometric function, and so on). Applying the chain rule under this thought repeatedly, any partial derivation can be broken down into easily differentiable functions.\par
Automatic differentiation has three merits: small amount of calculation, no theorical error, and no need to calculate the derivation manually. Recently, this technology is established as an indispensable element in DNN when optimizing (\ref{eq: 1}).

\subsection{Theoretical analysis for maximizing SDR}
\begin{figure}
\begin{center}
\includegraphics[width=6cm]{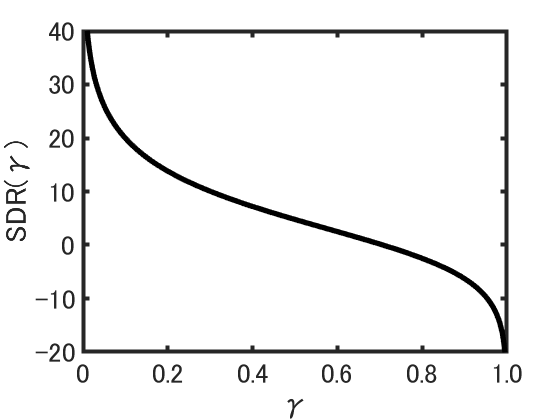}
\caption{Relation between $\gamma$ and SDR.}
\label{fig: 2}
\end{center}
\vspace{-12pt}
\end{figure}

SDR means how the estimated signal is similar to the clean signal. In this section, we simply show this fact in the following discussion. The $\bm{\hat{s}}$ consists of two components: $\bm{s}_{\rm target}$ and $\bm{s}_{\rm other}$. $\bm{s}_{\rm other}$ represents component in residual subspace for $\bm{s}_{\rm target}$, which is shown in 
\begin{align}
\bm{\hat{s}} = \sqrt{1-\gamma^2} \bm{s}_{\rm target} + \gamma \bm{s}_{\rm other},
\end{align}
where $0 \leq \gamma \leq 1$. Suppose $\|\bm{\hat{s}}\| = 1$, $\|\bm{s}_{\rm target}\| = 1$, $\|\bm{s}_{\rm other}\|=1$ and $\bm{s}_{\rm target} \perp \bm{s}_{\rm other}$, we obtain ${\rm SDR}(\gamma)$ as
\begin{align}
{\rm SDR}(\gamma) = 10 \log_{10} \frac{1-\gamma^2}{\gamma^2}.
\end{align}
Figure \ref{fig: 2} shows a relation between $\gamma$ and ${\rm SDR}(\gamma)$. If $\gamma \to 0$, $\bm{\hat{s}}$ is getting closer to $\bm{s}_{\rm target}$. This fact means that maximizing SDR leads $\bm{\hat{s}}$ to be in $\bm{s}_{\rm target}$. 

\section{Experiment}
\label{sec:exp}
\subsection{Experimental condition}
To evaluate the proposed method, we conduct a simple experiment. In this experiment, we use $s_t = \sin (12\pi t/T)$ as the clean signal. A noise signal is prepared as random variables with the uniform distribution. They are mixed with SNR of 10, 0 and -10 dB. $T=600$ and $\{ t \in \mathbb{Z} \mid 0 \leq t \leq 600\}$. \par
We adopt recurrent neural network (RNN) as DNN. RNN is a simple network structure developed to express time series. The size of input and output time series are both 100.\par
Training DNN is conducted as follows. First, we prepare the mixture signal, whose noise is different from that in the separation step, meanwhile clean signal is same. A training data contains the mixture signal and the clean signal. Second, we divide the training data by 100 samples shifting one sample at a time. Last, we optimize DNN using these datasets via stochastic gradient descent. A batch size is 50, an epoch is 500, and early stopping is used to judge convergence.\par
We use SDR and source-to-interference ratio (SIR)~\cite{vincent2006performance} as evaluation score implemented by~\cite{raffel2014mir_eval}. SIR indicates the ratio of the clean-signal part and the noise-signal part. Note that SIR do not take artificial distortion into account compared to SDR. SIR is defined as
\begin{align}
{\rm SIR} = 10 \log_{10} \frac{\|\bm{s}_{\rm target} \|^2}{\|\bm{e}_{\rm interf}\|^2},
\end{align}
where $\bm{e}_{\rm interf}$ is orthogonal projection of $\bm{\hat{s}}$ onto the subspace spanned by the noise signal and the delayed-noise signal. In addition, we compare L1 norm, L2 norm and the proposed method.

\subsection{Results}
Table \ref{tab:res1}, \ref{tab: res2} and \ref{tab: res3} are SDR and SIR of the conventional methods and the proposed method. It is obvious that the proposed method outperforms the conventional methods. This means that the proposed method truly reduces the noise signal compared to the conventional methods.

\begin{table}[t]
\begin{center}
  \caption{SDR [dB] and SIR [dB] when SNR of mixture signal is 10 dB}
   \scalebox{1}{
  \begin{tabular}{c|c c} \hline
    & SDR & SIR \\ \hline 
    L1 norm & 18.1 & 18.3 \\
    L2 norm & 18.3 & 19.2 \\
    Proposed & $\bm{24.8}$ & $\bm{25.0}$ \\ \hline
  \end{tabular}
}
  \label{tab:res1}
\end{center}
\begin{center}
  \caption{SDR [dB] and SIR [dB] when SNR of mixture signal is 0 dB}
   \scalebox{1}{
  \begin{tabular}{c|c c} \hline
    & SDR & SIR \\ \hline 
    L1 norm & 13.5 & 13.6 \\
    L2 norm & 14.5 & 14.8  \\
    Proposed & $\bm{17.7}$ & $\bm{18.0}$ \\ \hline
  \end{tabular}
}
  \label{tab: res2}
\end{center}
\begin{center}
  \caption{SDR [dB] and SIR [dB] when SNR of mixture signal is -10 dB}
   \scalebox{1}{
  \begin{tabular}{c|c c} \hline
    & SDR & SIR \\ \hline 
    L1 norm & 8.5 & 9.1 \\
    L2 norm & 9.2 & 9.3  \\
    Proposed & $\bm{10.9}$ & $\bm{11.6}$ \\ \hline
  \end{tabular}
}
  \label{tab: res3}
\end{center}
\end{table}

\begin{figure}
\includegraphics[width=8.5cm]{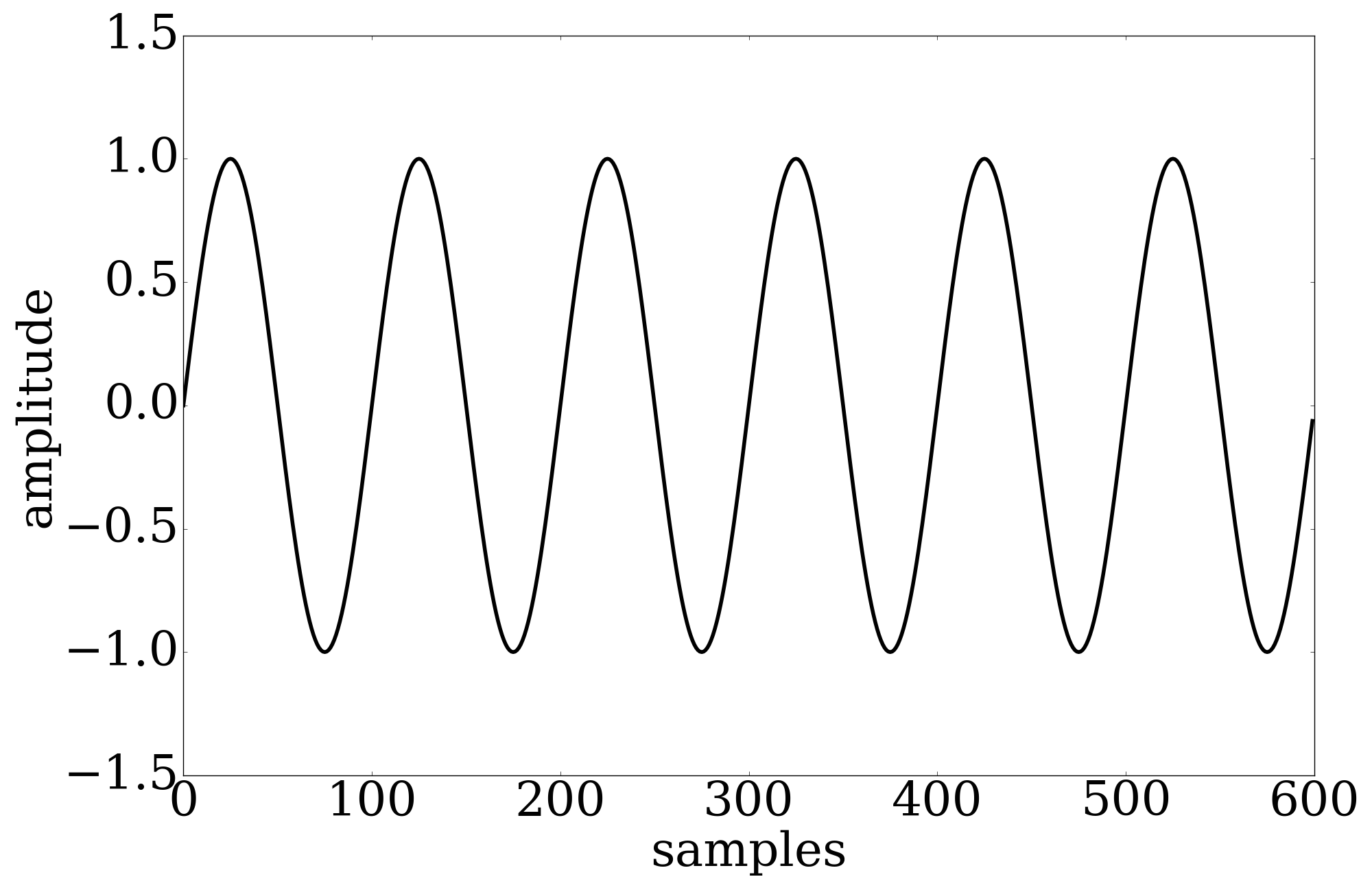}
\caption{Clean signal used in experiment when SNR is 10 dB.}
\label{fig: 3}
\end{figure}

\begin{figure}
\includegraphics[width=8.5cm]{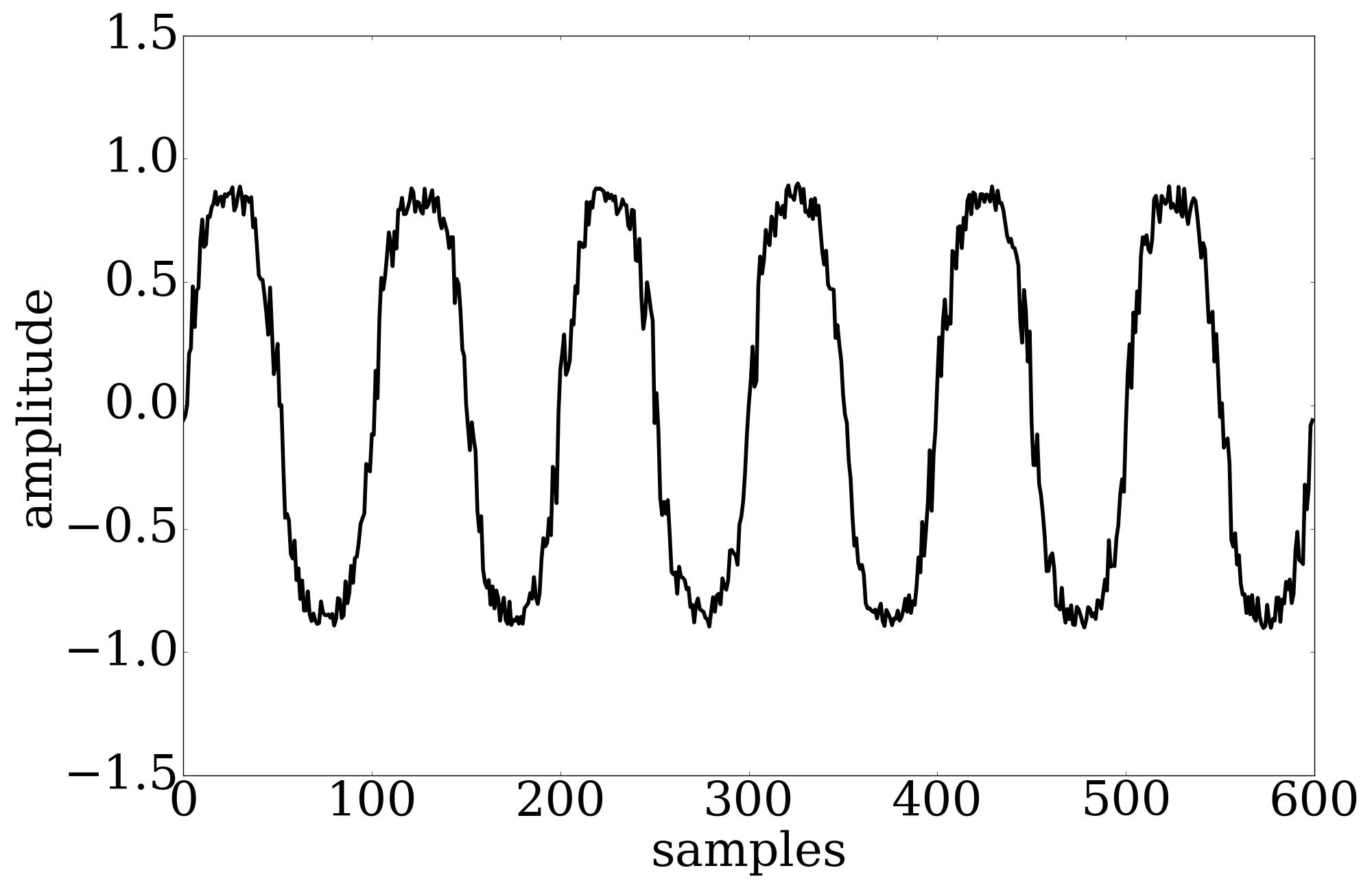}
\caption{Estimated signal by L2 norm when SNR is 10 dB.}
\label{fig: 4}
\end{figure}

\begin{figure}
\includegraphics[width=8.5cm]{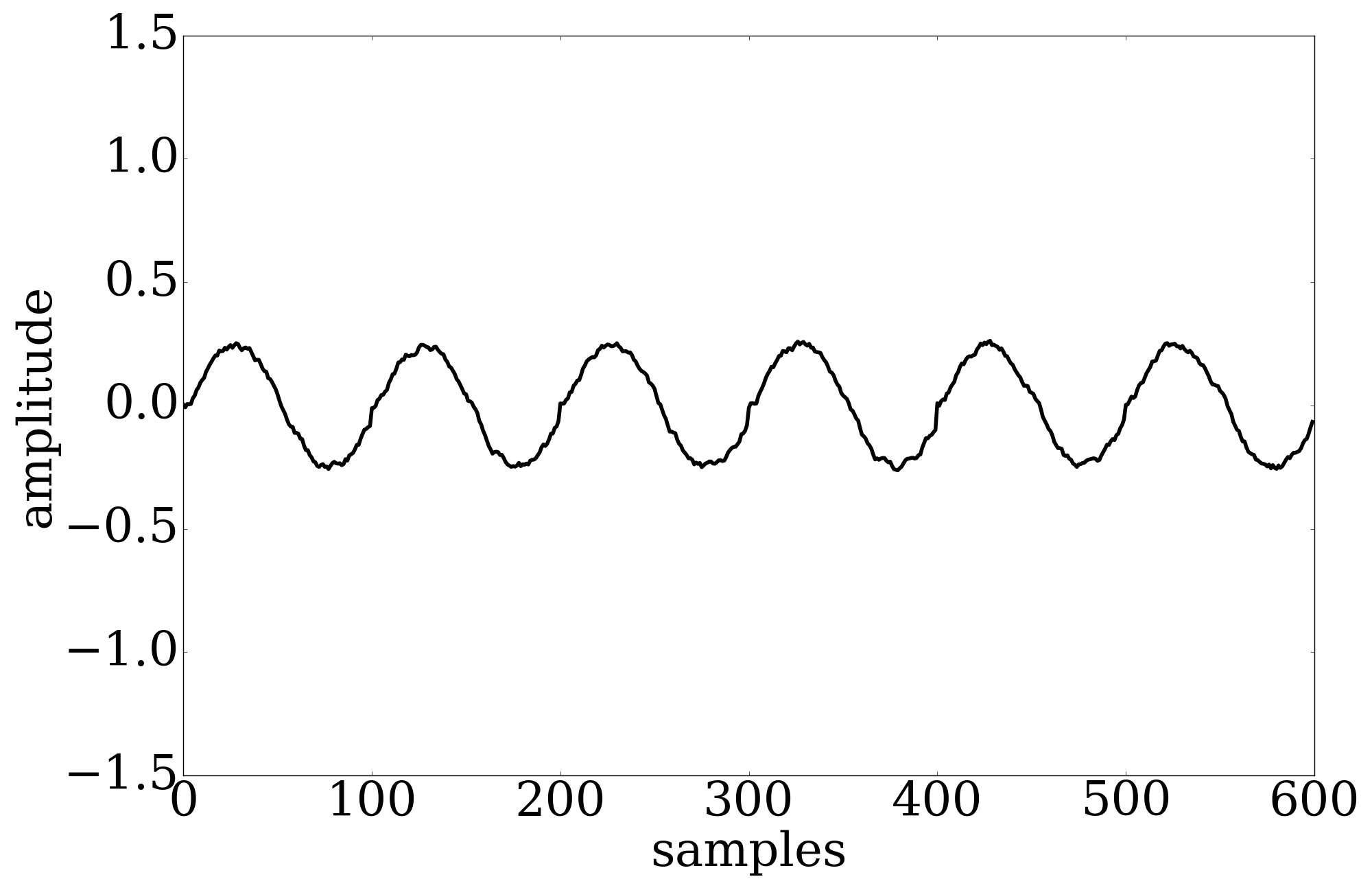}
\caption{Estimated signal by proposed method when SNR is 10 dB.}
\label{fig: 5}
\end{figure}

Figure \ref{fig: 3}, \ref{fig: 4} and \ref{fig: 5}  are the clean signal used in the experiment, an estimated signal based on L2 norm, and an estimated signal based on proposed method, respectively when SNR of the mixture signal is 10 dB. From these figures, it is confirmed that the proposed method reduces the noise signal obviously compared to the conventional method because jaggedness which indicate the noise signal is decreasing. On the other hand, amplitude of Fig. \ref{fig: 5} is smaller than that of Fig. \ref{fig: 3} and Fig. \ref{fig: 4}. It is because SDR does not care about amplitude reconstruction. For example, if $\hat{\bf s}$ is multiplied by two, $\bm{s}_{\rm target}$ doubles but SDR does not change. 
\vspace{-5pt}

\section{Conclusion}
\label{sec:concl}
In this paper, we proposed a DNN which is optimized by maximizing SDR. Conventional objective functions have a drawback that they only bring the estimated signal by DNN closer to the clean signal. The key idea of the proposed method is using SDR as the cost function, which enables us to reduce the noise signal component directly. From the experimental results, the proposed method outperforms the conventional methods in many cases. Future work should be conducted for more actual cases, for example vocal source separation, and so on.

\newpage

\bibliographystyle{IEEEbib}
\bibliography{refs}

\end{document}